\documentclass[pdflatex,sn-mathphys-num,iicol]{sn-jnl}
\pdfoutput=1

\usepackage[T1]{fontenc} 
\usepackage{subcaption}
\usepackage{xspace}
\usepackage{hyperref}
\usepackage{multirow}
\usepackage{amsmath}
\usepackage{listings}
\usepackage{xcolor}
\usepackage{fvextra}
\usepackage{orcidlink}
\usepackage{dblfloatfix}

\definecolor{yamlkey}{RGB}{0,0,139}       
\definecolor{yamlstring}{RGB}{160,32,240} 
\definecolor{yamlvalue}{RGB}{34,139,34}   
\definecolor{yamlcomment}{RGB}{128,128,128} 

\lstdefinelanguage{YAML}{
  keywords={true,false,null,y,n},
  keywordstyle=\color{yamlvalue}\bfseries,
  basicstyle=\ttfamily\small,
  sensitive=false,
  comment=[l]{\#},
  morecomment=[s]{/*}{*/},
  commentstyle=\color{yamlcomment}\ttfamily,
  stringstyle=\color{yamlstring}\ttfamily,
  moredelim=[l][\color{yamlkey}]{\ },
  moredelim=*[l][\color{yamlkey}]{\ Edgar}, 
  keepspaces=true,
  breaklines=true,
  frame=single,
  literate={:}{{\textcolor{red}{:}}}1 
}

\newcommand{\mycode}[1]{\EscVerb[breaklines, breakanywhere, fontfamily=tt]{#1}}

\newcommand*{\mt}{\ensuremath{m_{t}\xspace}}
\newcommand*{\Jpsi}{\ensuremath{J/\psi}\xspace}

\newcommand*{\electronvolt}{\text{e\kern-0.1em V}}

\newcommand*{\GeV}{\ensuremath{\text{G\electronvolt}}\xspace}

\graphicspath{{./}}

\begin{document}

\title{Combinations of measurements that are simultaneous fits of parameters of interest and systematic uncertainties using the BLUE method}

\author[a]{Tomas Dado \orcidlink{0000-0002-7050-2669}}\email{tomas.dado@cern.ch}
\author[b]{Mark Owen \orcidlink{0000-0001-6820-0488}}\email{mark.owen@glasgow.ac.uk}
\author[c]{Michele Pinamonti \orcidlink{0000-0002-5282-5050}}\email{michele.pinamonti@ts.infn.it}

\affil[a]{CERN, Esplanade des Particules 1, 1217 Meyrin, Switzerland}
\affil[b]{School of Physics \& Astronomy, The University of Glasgow, University Avenue, Glasgow, G12 8QQ, UK}
\affil[c]{INFN Gruppo Collegato di Udine, Sezione di Trieste, Via delle Scienze 208, 33100 Udine, Italy}

\abstract{Combining estimates of the same physics parameter obtained from different measurements improves the precision and robustness of the parameter determination.
Modern particle physics measurements
are often performed using likelihood fits that include the physics parameter(s) of interest together with nuisance parameters representing systematic uncertainties.
In high-statistics analyses, typical of many analyses at the Large Hadron Collider, these nuisance parameters can be constrained in the likelihood fits.
We describe how the Best Linear Unbiased Estimator method for combinations can be applied to both the estimates of the parameters of interest and the estimates of the nuisance parameters.
We show with concrete example combinations that including the nuisance parameters can improve the precision on the parameters of interest.
We show with pseudo-experiments that the uncertainty reported by the combination is reliable and that the approximate likelihood combination proposed in a previous publication and implemented in the Convino software reports an underestimated
uncertainty.
The method is implemented in an open-source software tool, Combiner.}

\maketitle
\flushbottom

\section{Introduction}
\label{sec-intro}

Measurements of physics parameters, such as the Higgs boson mass, Higgs boson couplings, the top quark mass, the strong coupling constant
and the parameters of the neutrino mixing matrix
are at the heart of experimental particle physics.
The existence of multiple experiments, for example, at the Large Hadron Collider~(LHC)~\cite{LHC} and in the neutrino sector, motivates the need to combine measurements
from different experiments and from different measurements within single experiments.
One technique for performing such combinations is the Best Linear Unbiased Estimate (BLUE)~\cite{Lyons:1988rp,Valassi:2003mu}, which requires the covariance matrix for the measurements
being combined.
This technique has been widely used in analyses where the physics parameters of interest (POIs) were extracted from a likelihood
fit in which only the POIs are free parameters (referred to as a statistical-only likelihood fit in the remainder of the paper); see, e.g., Refs.~\cite{lhcmtop,CMS:2020ezf,LHCb:2019epo,ATLAS:2019hhu}.
Systematic uncertainties are typically evaluated by repeating the fits with varied models
that represent the systematic uncertainties~\cite{vanDyk:2023tqz}. The systematic uncertainties can then be added in quadrature with the statistical uncertainty
from the fit to give the total uncertainty. Systematic uncertainties are normally assumed to be uncorrelated with each other, and
the covariance between any two measurements is then straightforward to estimate once the correlation between systematics in different measurements is established.

A feature of the LHC analysis era is that many measurements of physics parameters are extracted from likelihood fits that include both the physics parameters of interest and a set of parameters that represent the systematic
uncertainties, often referred to as nuisance parameters (NP).
The results of such fits (referred to as full-likelihood fits in the following) are estimates of both the POIs and the NPs together with the corresponding covariance matrix.
As well as conveniently including the systematic uncertainties into the likelihood, this means the data can add to the knowledge on the NPs and reduce
the uncertainty in the POIs.
Recent examples include measurements of the $t\bar{t}$~cross-section near threshold~\cite{CMS:2025kzt,ATLAS:2026dbe}.
If the full likelihood is available, such measurements can conveniently be combined by forming the joint likelihood of the measurements (see, e.g., Ref.~\cite{ATLAS:2026pdi}); however,
in many cases the full-likelihoods are not publicly available and hence combination techniques starting from the results of the fits are needed.
The NPs after a full-likelihood fit will typically have non-zero correlation, which at first glance appears to break the view
of systematic uncertainties as being uncorrelated with each other.
Ref.~\cite{Kieseler:2017kxl} proposes an approximate method and tool (Convino) for combining such measurements, indicating that the BLUE method
cannot be used without information beyond the covariance matrix and central values of the full-likelihood fits.
This method has subsequently been used in several recent publications~\cite{ATLAS:2022aof,CMS:2021oxn,CMS:2020cso,CMS:2019oeb}.
While the latter three combine statistical-only likelihood fits (in which case the Convino method corresponds to BLUE), Ref.~\cite{ATLAS:2022aof}
combines a CMS full-likelihood fit with an ATLAS statistical-only likelihood fit.
Ref.~\cite{matrixPLpaper} details how the contributions of the systematic uncertainties to measurements from full-likelihood fits can be obtained
from the covariance matrix if the measurement is in the Gaussian regime\footnote{For measurements not in the Gaussian regime, the contributions of the systematic uncertainties can be obtained from varying the global observables (see Section~\ref{sec-example1} for the definition of the global observables).}.
The BLUE method can then be used to combine such measurements based on the normal assumption that the systematic uncertainties
are uncorrelated with each other.
The authors of Ref.~\cite{matrixPLpaper} noted that it is possible to combine either the POIs or the POIs and the NPs from full-likelihood measurements,
but showed for the example given in the paper that the two options gave the same precision.
In this paper, we show that the combination of POIs and NPs can, in certain situations, result in a smaller uncertainty for the combined POIs
than when only the POIs are combined.
We demonstrate that this behaviour occurs even if only one of the input measurements is a full-likelihood measurement.
Two examples are investigated: one is purely hypothetical, and the other combines two recent ATLAS top quark mass measurements.
For the first example, we also perform a combination with the Convino tool~\cite{Kieseler:2017kxl} and show with pseudo-data that the
uncertainty from this method is underestimated.
The paper is accompanied by a software tool called Combiner~\cite{combiner}, which implements the logic described in the paper.
The tool can be used to combine any number of analyses that are based on either statistical-only or full-likelihood fits.

The paper is structured as follows: in Section~\ref{sec-mathSetup} we review the mathematical framework for combining multiple full-likelihood analyses.
In Section~\ref{sec-nonprof}, we examine how statistical-only likelihood fits can be incorporated into the setup
and show that the NP measurements from a full-likelihood analysis can improve the combined estimate of POIs even when one full-likelihood
analysis is combined with one statistical-only likelihood analysis.
In Section~\ref{sec-example1}, we set up two hypothetical analyses using full-likelihood fits.
The analyses are combined with BLUE using the POIs only and using the POIs and NPs. The results are compared with a combination using Convino
and are validated using pseudo-experiments. In Section~\ref{sec-example2}, we combine two recent ATLAS top quark mass measurements and
show that including the NPs reduces the uncertainty on the combined top quark mass estimate.
We finally summarise the paper in Section~\ref{sec-summary}.

\section{Mathematical framework}
\label{sec-mathSetup}

The output of a full-likelihood analysis that is in the Gaussian regime is $M$~estimates of the POIs and $N$~estimates of the NPs, which we collectively
denote as $\vec{\theta}$~where the
best fit values for the parameters $\vec{\hat{\theta}}$~are complemented with the covariance matrix $C$~that encodes the uncertainties
on the parameters and the correlations between them. To combine this fit with another fit where at least some of the POIs and NPs are in common,
we need to estimate the correlation between the estimates from each analysis.
For many cases, the two analyses will have been performed on orthogonal datasets (e.g. analyses from different experiments or analyses within the
same experiment that use selections that ensure orthogonality). In these cases, the correlation between the fitted parameters arises from the common
sources of systematic uncertainties.
Following~\cite{matrixPLpaper}, the contribution to the uncertainty on each parameter from every source of systematic uncertainty can often be estimated
from the covariance matrix $C$~and we denote the contribution of uncertainty source $k$~on the estimate of parameter~$i$~as $\Gamma_{ik}$.
The covariance between two analyses $A$~and $B$~can then be written as:
\begin{align}
C_{AB} =
\begin{pmatrix}
C_A  &  D \\
D  & C_B
\end{pmatrix}
\end{align}
where the elements of the cross-correlation matrix are $D_{ij} = \Gamma_{ik}\Gamma_{jk}$~\footnote{In the paper, unless otherwise stated, we use the Einstein summation convention where repeated indices denote a sum over that index.}~and $C_A$~and $C_B$~are the covariance matrices for the two analyses $A$~and $B$.
In the case that the analyses $A$~and $B$~have some statistical correlation (for example they use overlapping sets of events), the statistical covariance can be estimated
using pseudo-experiments on the individual analyses to produce a statistical covariance matrix $S$~and the cross-correlation matrix must then include
those cross-correlations with $D_{ij} = \Gamma_{ik}\Gamma_{jk} + S_{ij}$~where $S_{ij}$~is the element of $S$~that expresses the statistical covariance between the estimates
$i$~and $j$.
As we are dealing with Gaussian measurements, the best-fit values for the parameters are obtained by minimising the $\chi^2$~
\begin{align}
\chi^2 = 
\left( \vec{\hat{\theta}} - \vec{\theta} \right)^T
C_{AB}^{-1}
\left( \vec{\hat{\theta}} - \vec{\theta} \right)
\end{align}
where the vector $\vec{\hat{\theta}}$~contains the best-fit values for all the parameters (both POIs and NPs) in analyses $A$~and $B$.
The minimisation of this $\chi^2$~can be done analytically and corresponds to the well-known BLUE method~\cite{Lyons:1988rp,Valassi:2003mu}.
The combined estimate for a parameter $\theta_{\alpha}$~is
\begin{equation}
\label{eqn-blue-est}
\theta_{\alpha} = (U^T C_{AB}^{-1} U)^{-1}_{\alpha j} (U^T C_{AB}^{-1})_{j i} \hat{\theta}_i
\end{equation}
where the $U$~matrix is defined such that $U_{i\alpha}=1$~if $i$~is an estimate of parameter $\alpha$~and zero otherwise.
As discussed in Ref.~\cite{Valassi:2003mu}, the combined estimate of each parameter receives contributions from all parameters that are estimated by both $A$~and $B$.
The weight of an estimate $\hat{\theta}_i$~on the combined estimate $\theta_{\alpha}$~is $\lambda_{\alpha i} = (U^T C_{AB}^{-1} U)^{-1}_{\alpha j} (U^T C_{AB}^{-1})_{j i}$~and
is referred to as the BLUE weight. These weights obey the constraints:
\begin{align}
\sum_{i}^{\text{estimates of~}\alpha} \lambda_{\alpha i}  = 1 \\
\sum_{i}^{\text{estimates not of~}\alpha} \lambda_{\alpha i}  = 0.
\end{align}
The estimates of the NPs from the two analyses $A$~and $B$~will therefore contribute to the combined estimates of the POIs and the covariance matrix of the combined estimates is:
\begin{align}
c_{\alpha\beta} = (U^T C_{AB}^{-1} U)^{-1}_{\alpha\beta}.\label{eq-combCov}
\end{align}

It is a choice to include the estimates of the NPs in the above procedure; it is also possible to choose to consider only the POIs from the two analyses. The mathematics
remains the same (as it makes no distinction between POIs and NPs), and the covariance matrix $C_{AB}$~is reduced to $2M\times 2M$, rather than $(2M+2N) \times (2M+2N)$.
The resulting combined estimates will only have contributions from the POIs.
As this reduced combination uses less data, it follows from standard information theory~\cite{infoTheoryBook} that the uncertainty on the POIs obtained from combining both the POIs and the NPs is less than or equal to
the uncertainty on the POIs obtained from combining only the POIs. We examine specific examples of this situation in Section~\ref{sec-example1}.

\section{Combining non-profiled analyses}
\label{sec-nonprof}

The framework described above imagines analyses that have performed a full-likelihood fit to obtain estimates of the POIs and the NPs.
There are many measurements where the fit to the data uses only the statistical uncertainties and the systematic uncertainties
are propagated through the measurement.
The outputs of such a measurement are the best-fit value of the POIs,
the covariance between them ($O$)~and the contribution of each systematic uncertainty $k$~to each POI $i$, $\Gamma_{ik}$.
The prior measurements of the NPs can be thought of as having a best-fit value of zero and uncertainty one that is uncorrelated with all other NPs.
This allows us to enlarge the covariance matrix of a measurement using a statistical-only likelihood as follows:
\begin{equation}
C=
\begin{pmatrix}
O & \Gamma \\
\Gamma^T & I
\end{pmatrix}
\label{eq-cov-np}
\end{equation}
where $I$~is the identity matrix. With this matrix in hand, an analysis using a statistical-only likelihood can be included in exactly the same way as an analysis using a full-likelihood in
the combined estimates discussed in Section~\ref{sec-mathSetup}.

We now apply this setup to a full-likelihood analysis $A$~and a statistical-only likelihood analysis $B$, each of which has one POI $p$~and
$N$~NPs $\vec{n}$.
Denoting the two estimates of the POI as $p_A$~and $p_B$, the estimates of the NPs from analysis $A$~as $\vec{n_{A}}$~and the prior
estimates of the NPs implicit in analysis $B$~as $\vec{n_{B}}=\vec{0}$,
the combined estimate of the POI $p$~obtained with Equation~\ref{eqn-blue-est} is:
\begin{align}
p &= \lambda_{p_A} p_A + \lambda_{p_B} p_B + \sum_{i=0}^N\left(\lambda_{n_{A,i}} n_{A,i} + \lambda_{n_{B,i}} n_{B,i} \right)\nonumber \\
 &= \lambda_{p_A} p_A + (1 - \lambda_{p_A}) p_B + \sum_{i=0}^N \lambda_{n_{A,i}} n_{A,i}
\end{align}
where $\lambda_{p_j}$~are the BLUE weights for the POIs from measurement $j$~and $\lambda_{n_{j,i}}$~are the BLUE weights for NP $i$~from
measurement $j$.
We observe that the combined estimate now comprises the estimates of the two POIs plus a contribution from the estimates of the
NPs from measurement $A$. One way to view this is that the estimates of the NPs from analysis $A$~are being used to improve the
estimate of the POI from analysis $B$. In this way, the combined estimate of the POI has a smaller variance than if one were to combine
only the two POIs from the two analyses. This behaviour is demonstrated in the combinations described in Sections~\ref{sec-example1}~and~\ref{sec-example2}.
Appendix~\ref{app-combiner} provides a description of the configuration files for the Combiner tool.

\section{Example combination 1}
\label{sec-example1}

To illustrate the features of the combined estimates discussed in Section~\ref{sec-mathSetup}, we consider two example analyses that are to be combined.
The two analyses both seek to measure the rate of a signal process, $\mu$, where $\mu=1$~is defined to be the nominal prediction for the signal.
There is a background process which is subject to systematic uncertainties.
Both analyses use a single signal region (SR) which contains both signal and background and two control regions (CR) which contain only background events.
A full-likelihood fit is performed in each analysis to the signal and control regions, where the events in the control regions provide information
on the NPs, $\theta_1$~and $\theta_2$. The likelihood in each analysis is constructed as:
\begin{align}
\mathcal{L}(d | \mu, \theta_1, \theta_2) = P\left(d_{SR} | s(\mu) + b_{SR}(\theta_1, \theta_2)\right) \nonumber \\
\times \prod_{i}^{CR} P\left(d_{CRi} | b_{CRi}(\theta_1, \theta_2)\right) \prod_s G(g_s | \theta_s, \sigma=1)
\end{align}
where $d_{SR}$~and $d_{CRi}$~are the numbers of data events in the signal region and the two control regions ($i=1,2$), $P$~denotes a Poisson probability distribution function (PDF), and $G$~denotes a Gaussian PDF. The number of expected signal events
in the SR is $s(\mu)=s_0 \mu$, where $s_0$~is the nominal expectation for the signal. The numbers of expected background events in the SR and CRs are $b_{SR}$~and $b_{CRi}$. The final term contains the Gaussian
constraints on the systematic uncertainties, where $g_s$~are referred to as the global observables and have a nominal value of zero. The Asimov
dataset~\cite{Cowan:2010js} is defined to be the expected number of events for $\mu=1$~and $g_s=0$.

\begin{figure}[htbp]
\centering
\begin{subfigure}{0.45\textwidth}
\includegraphics[width=\textwidth]{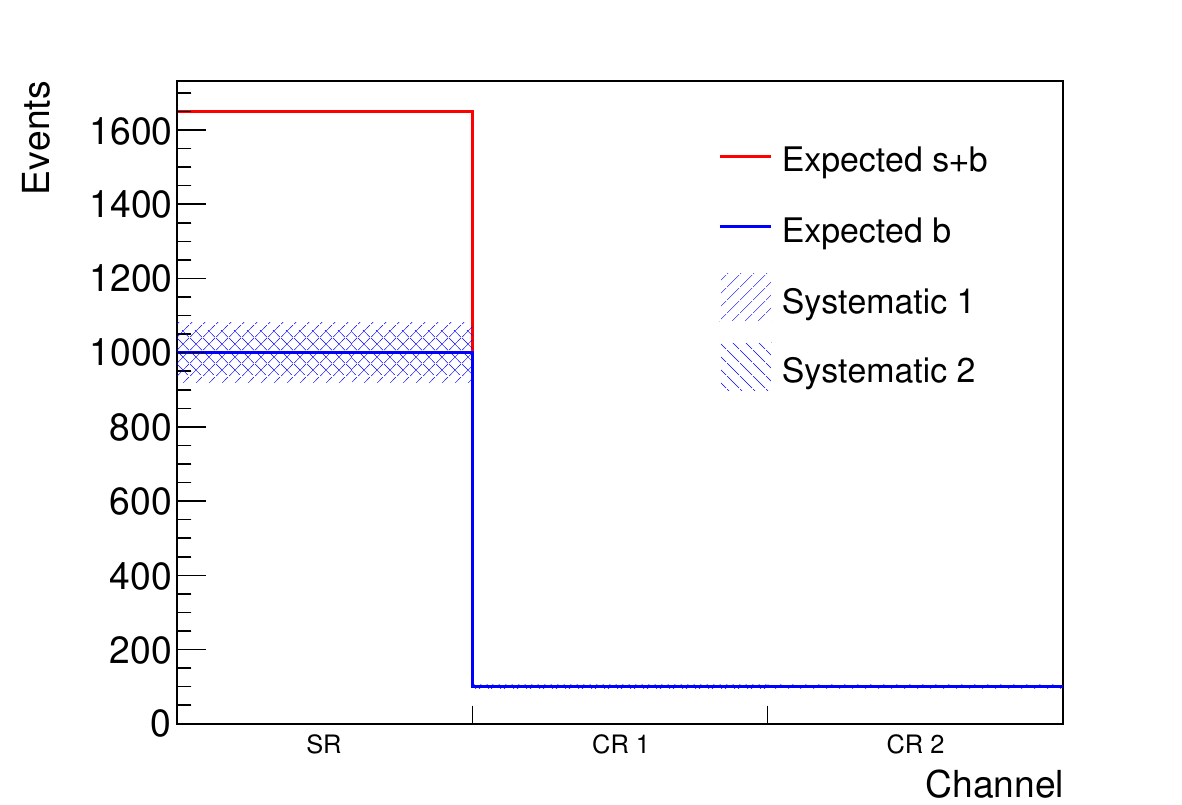}
\caption{}
\end{subfigure}
\begin{subfigure}{0.45\textwidth}
\includegraphics[width=\textwidth]{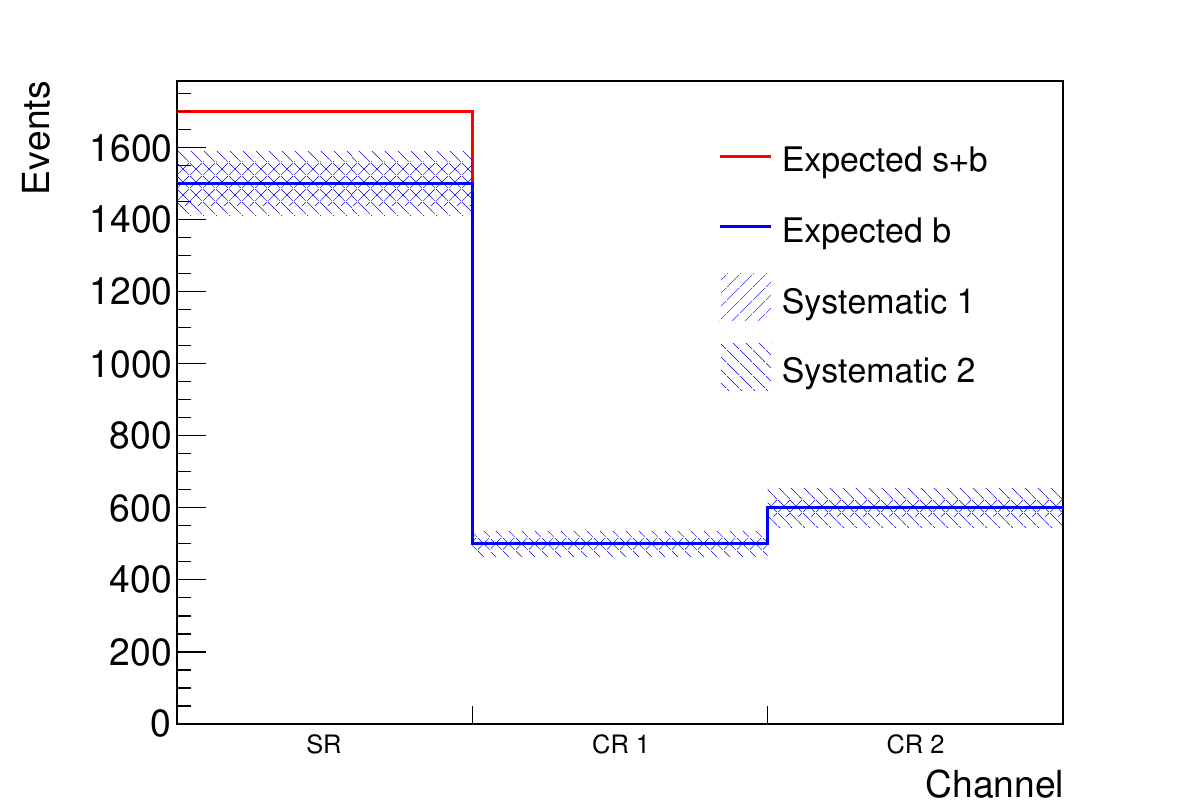}
\caption{}
\end{subfigure}
\caption{
The expected numbers of background events~(blue) and signal plus background~(red) are shown for (a) analysis $A$~and (b) analysis $B$. The sizes of the hashed areas
show the uncertainties in the expected number of background events that originate from the systematic uncertainties. The signal distribution assumes $\mu=1$.
}
\label{fig-ex-chanPlots}
\end{figure}

Analysis $A$~is set up such that it has high sensitivity to $\mu$~but low constraining power for the NP. Analysis $B$~is set up such that it has low sensitivity
to $\mu$~but high constraining power on the NPs.
The expected signal and background events for the two analyses are shown in Figure~\ref{fig-ex-chanPlots}.
In order to test a combination of a measurement using a statistical-only fit, the fit for analysis $A$~is also performed with only the SR. In this configuration, there is one data bin
and one free parameter ($\mu$) and so the NPs cannot be constrained. The covariance matrix from this fit has exactly the form shown in Equation~\ref{eq-cov-np}, allowing
for a test where analysis $A$~is effectively a statistical-only likelihood.

The uncertainties obtained from the fits to the Asimov dataset for the two analyses are shown in Table~\ref{tab-ex-res}. Analysis $A$~has significantly better precision on the POI
than analysis $B$, while analysis $B$~has a more precise estimate of systematic 2. The combination using only the POIs makes only a mild improvement
on the result from analysis $A$, due to the relatively poor precision of analysis $B$. Including the estimates of the NPs in the combination
significantly improves the precision of the POI. This is an example in which performing the combination across the full set of estimated parameters
improves on the combination of only the POIs. The origin of this improvement can be understood by examining the BLUE weights of the combinations,
which are shown in Table~\ref{tab-ex-weights}. In the combination of the POIs and NPs, the estimates of the second NP carry substantial weight in the combined
estimate of the POI and the precision of this estimate is improved by including the estimates of the NPs.
Table~\ref{tab-ex-res}~also shows the expected uncertainty for a combination using results from a fit on only the SR from analysis $A$~and the SR+CR from analysis $B$.
This corresponds to combining a statistical-only fit with a full-likelihood fit and the corresponding BLUE weights are shown in Table~\ref{tab-ex-weights}.
As anticipated in Section~\ref{sec-nonprof}, the precision on the combined POI is improved by including the NP estimates in the combination.

\begin{table*}[htbp]
\centering
\begin{tabular}{l | c c c}
Measurement & POI & Syst1 & Syst2 \\ \hline 
Analysis $A$ & 0.13 & 0.83 & 0.91 \\ 
Analysis $B$ & 0.28 & 0.92 & 0.50 \\ 
Analysis $A$~SR only & 0.17 & 1.00 & 1.00 \\ \hline 
Combination of analysis $A$~\& $B$, using POI \& NPs & 0.10 & 0.82 & 0.47 \\ 
Combination of analysis $A$~\& $B$, using only POI & 0.12 & - & - \\ 
Combination of analysis $A$~SR \& analysis $B$, using POI \& NPs & 0.11 & 0.92 & 0.48 \\ 
Combination of analysis $A$~SR \& analysis $B$, using only POI & 0.16 & - & - \\ 
\end{tabular}
\caption{Uncertainties obtained from fits to the two example analyses and the uncertainties obtained by combining the two analyses. The third line refers to the fit using only the SR of Analysis $A$, and neglecting the CR. The first combination uses the estimates of both the POIs and NPs from both analyses, while the second combination uses only the estimates of the POI from the two analyses. The last two combinations use only the SR from analysis $A$, using either the estimates of both the POIs and NPs or just the estimates of POIs.}
\label{tab-ex-res}
\end{table*}

\begin{table*}
\centering
\resizebox{\textwidth}{!}{%
\begin{tabular}{l | c c c c c c c}
 Combination & POI 1 & POI 2 & NP 1,1 & NP 1,2 & NP 2,1 & NP 2,2 \\ \hline 
Analysis $A$~\& $B$, POI and NPs & 1.02 & -0.02 & 0.03 & -0.03 & 0.10 & -0.10 \\ 
Analysis $A$~\& $B$, POI only & 0.89 & 0.11 & - & - & - & - \\ 
Analysis $A$~SR \& analysis $B$, POI and NPs & 1.05 & -0.05 & 0.10 & -0.10 & 0.11 & -0.11 \\ 
Analysis $A$~SR \& analysis $B$, POI only & 1.05 & -0.05 & - & - & - & - \\ 
\end{tabular}
}
\caption{BLUE weights for the combinations of the example analyses. The columns POI 1 and POI 2 show the weights for the estimates of the POI from analysis 1 and 2. The columns NP $i,j$~show the weight of the estimate of NP $i$~from analysis $j$.}
\label{tab-ex-weights}
\end{table*}

\begin{figure*}[htbp]
\centering
\begin{subfigure}{0.45\textwidth}
\includegraphics[width=\textwidth]{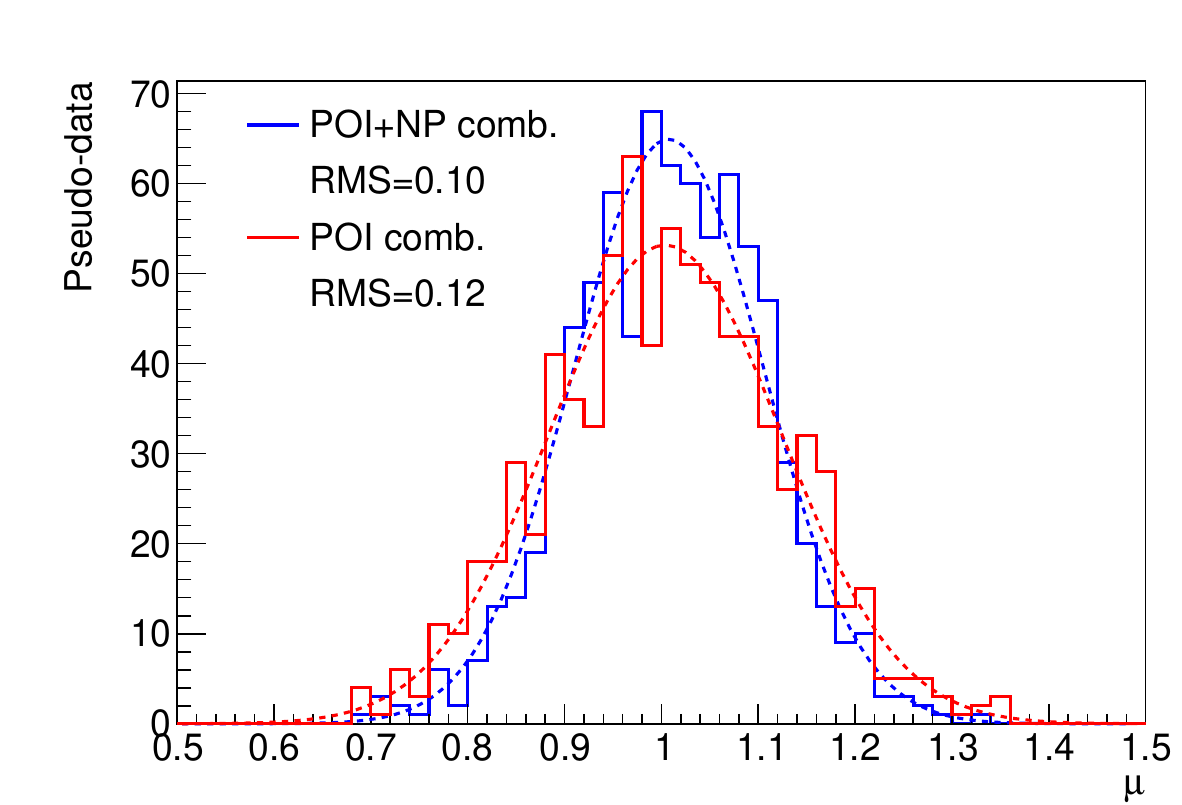}
\caption{}
\end{subfigure}
\begin{subfigure}{0.45\textwidth}
\includegraphics[width=\textwidth]{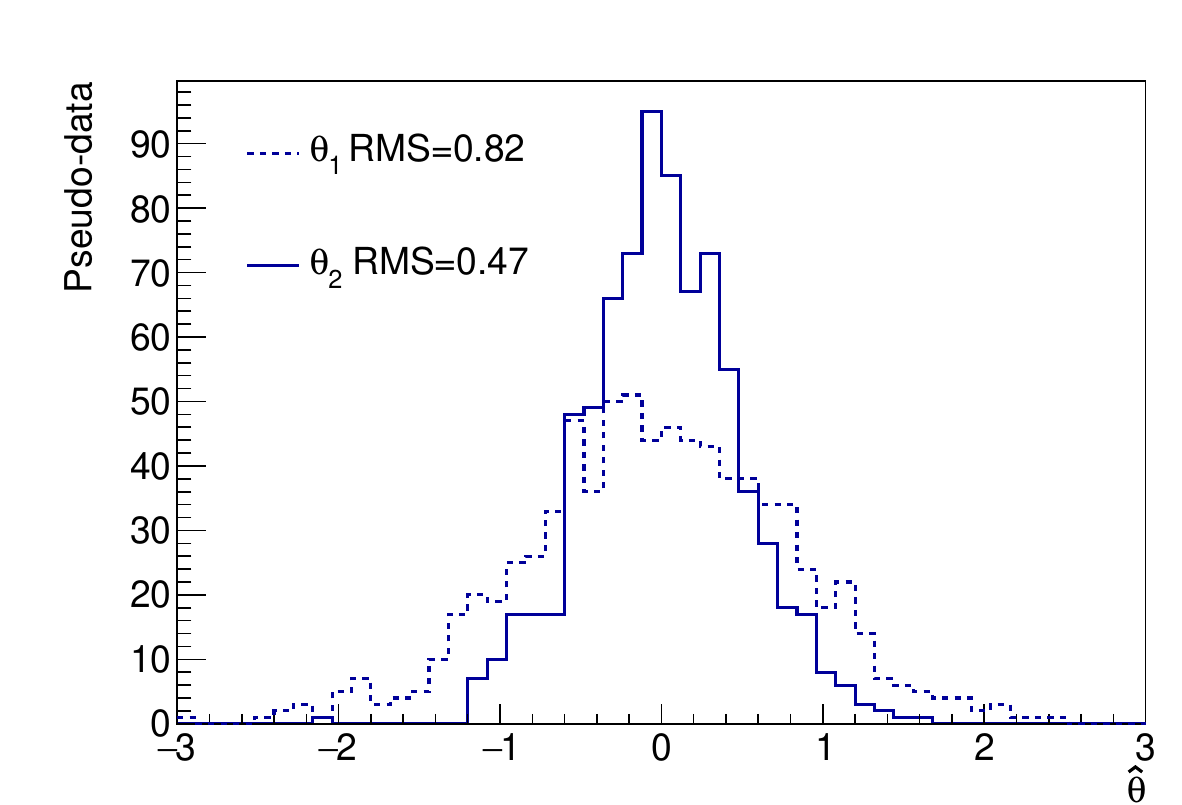}
\caption{}
\end{subfigure}
\caption{
The fitted $\mu$~values from the pseudo-data for the two combinations of analyses $A$~and $B$~are shown in (a). The solid blue line shows the combinations using the estimates of the POI and the NPs and
the solid red line shows the combination using the estimates of only the POIs. The dashed lines show Gaussian distributions that have been fitted to the pseudo-data results. The fitted values for the NPs ($\hat{\theta}$) from the combinations of the estimates of the POI and NPs are shown in (b), the dashed line shows
NP1 and the solid line shows NP2. In all cases the RMS of the pseudo-data is shown and agrees with the expected uncertainty shown in Table~\ref{tab-ex-res}. The relative uncertainty on the RMS from the limited number of pseudo-data is less than $3\%$ in all cases.
}
\label{fig-ex1-toys}
\end{figure*}

The validity of the combinations is tested by performing pseudo-experiments on the analyses. The global observables $g_s$~are fluctuated according to Gaussian distributions with mean zero and width one. For each pseudo-experiment, the values of $g_s$~are common to both analyses (representing correlated systematic uncertainties).
In each analysis, the number of events in each signal and control region is fluctuated according to a Poisson distribution. In each pseudo-experiment, the fit
for each analysis is performed and the outputs of the fits are then used to perform the same two combinations performed above on the Asimov datasets.
The means of the combined POIs (NPs) are found to be consistent with one (zero) and the RMS agrees with the expected precision shown in Table~\ref{tab-ex-res} for all combinations.
The results from the pseudo-experiments for the two combinations of analyses $A$~and $B$~are shown in Figure~\ref{fig-ex1-toys}.
This validates that the combination procedure is unbiased and has a reliable uncertainty estimate.

\subsection{Comparison with Convino}
\label{sec-example1-convino}

The Convino method presented in~\cite{Kieseler:2017kxl} presents an alternative method for combining full-likelihood analyses. It is notable that although this method uses the
full covariance matrix from each input analysis and the estimates of the POIs, the best-fit estimates for the NPs from each analysis are not used.
The combination of analyses $A$~and $B$~is performed with the Convino method and the method presented here. 
Pseudo-experiments are performed as described in the previous section and the RMS of the combined $\mu$ from the pseudo-experiments is shown in Table~\ref{tab-convino}.
While the uncertainties from the BLUE method match the RMS from the pseudo-experiments, the uncertainty from Convino is found to be underestimated compared to the RMS
from the pseudo-experiments. It is expected that this is because Convino is not using the best-fit estimates for the NPs in the approximate likelihood.
This behaviour is discussed in more detail in Appendix~\ref{app-convino}.

\begin{table*}
\centering
\begin{tabular}{l | c | c }
Combination method & Reported uncertainty of combination & RMS of combined $\mu$ \\ \hline
BLUE with POI only & 0.12 & 0.12 \\
BLUE with POI and NP & 0.10 & 0.10 \\
Convino  & 0.10 & 0.12 \\
\end{tabular}
\caption{Comparison of the uncertainty from different combination methods and the RMS of the combined $\mu$~derived from pseudo-experiments when combining analyses $A$~and $B$. The relative uncertainty on the RMS from the limited number of pseudo-data is less than $3\%$~in all cases.}
\label{tab-convino}
\end{table*}

\section{Example combination 2}
\label{sec-example2}

The example combination in the previous section serves to illustrate the features of combining both the POIs and NPs from full-likelihood analyses,
but it uses analyses that are significantly simpler than realistic measurements. To test the method in a more realistic environment, we use two recently published
ATLAS measurements of the top quark mass~($\mt$).
The first measurement uses $t\bar{t}$~events with a high transverse momentum (boosted) top quark~\cite{boosted}.
This analysis is a full-likelihood fit with non-negligible constraints on the NPs.
The likelihood and post-fit covariance matrix are publicly available in HepData~\cite{hepdata.158358}~and the latter provides the first input for the second example combination.
The second input for this example combination is the ATLAS measurement of the top quark mass using $t\bar{t}$~events where a $\Jpsi$~is produced from one of the top quark decays~\cite{jpsi}.
This analysis uses a statistical-only likelihood but shares several of the uncertainties in the boosted measurement and so the combination can explore the properties discussed in Section~\ref{sec-nonprof}.
The full breakdown of uncertainties is stored in HepData~\cite{hepdata.167264} and is used to perform the combination.
The $\Jpsi$~analysis has a large systematic uncertainty originating from the choice of recoil model in the $t\bar{t}$~Monte Carlo simulation. The NP for this uncertainty is constrained
in the boosted analysis and hence we expect that including the measurements of this NP into the combination will improve the overall precision on $\mt$.

\begin{table*}
\centering
\begin{tabular}{l | c c c c | c}
 \multirow{3}{*}{Combination} & \multicolumn{4}{c|}{BLUE weights} & \multirow{3}{*}{$\Delta \mt$ (\GeV)}\\
  &  \mt & \mt& Recoil NP  & Recoil NP \\ 
   & boosted & \Jpsi & boosted & \Jpsi \\ \hline 
POI only & 0.90 & 0.10 & - & -  & 0.50 \\ 
POI and recoil NP & 0.84 & 0.16 & -0.17 & 0.17 & 0.48 \\ 
\end{tabular}
\caption{BLUE weights and combined uncertainty in \mt~($\Delta \mt$) for the \mt~combinations of the ATLAS boosted and \Jpsi~measurements.}
\label{tab-mtop}
\end{table*}

To perform this combination, it is necessary to decide whether each of the systematic uncertainties in the two measurements has an equivalent uncertainty in the other measurement and is therefore correlated.
A fully accurate treatment of the correlations requires dedicated efforts by the ATLAS collaboration (similar to those performed for, e.g., Ref.~\cite{lhcmtop}).
In this paper, we aim only to have a sufficiently reasonable correlation model so that
we can explore the statistical features of the combination discussed in Section~\ref{sec-nonprof}.
As the two measurements use the same ATLAS dataset, the detector uncertainties largely have a clear mapping - uncertainties that have the same name are assumed to be equivalent.
One exception is the jet energy scale (JES) flavour uncertainties, where the boosted analysis uses a more detailed procedure.
The $b$-JES uncertainties in the $\Jpsi$~analysis are assumed to be uncorrelated with the three heavy-flavour JES uncertainties in the boosted analysis. This choice has negligible impact
on the combination as the JES uncertainties are very small in the $\Jpsi$~analysis.
Similarly, many of the $t\bar{t}$~and single top modelling uncertainties are derived from the same procedures and in these cases we also assume the uncertainties to be correlated
between the two measurements.
This includes the systematic uncertainty for the recoil model, which is the largest single uncertainty in the $\Jpsi$~analysis. 
The parton-shower and hadronisation uncertainty is decorrelated between the three observables used in the fit for the boosted measurement. We chose to correlate the uncertainty
in the \mt~sensitive variable with the corresponding uncertainty in the $\Jpsi$~analysis. 
Uncertainties in the background estimates for the two analyses are considered to be uncorrelated due to the very different phase spaces targeted by the measurements.
Any systematic uncertainties that appear in only one of the measurements (e.g. the soft-muon systematic uncertainties in the $\Jpsi$~measurement) are assumed to be uncorrelated
with all uncertainties in the second measurement.
The full set of correlation assumptions is encoded in the configuration files described in Appendix~\ref{app-combiner}.
The overall correlation between the \mt~estimators for the boosted and $\Jpsi$~analyses is found to be $0.01$.

The combination is first performed using only the two POI measurements of \mt. This yields a combined measurement of $\mt$~with an uncertainty of $0.50$~\GeV.
The BLUE weights are shown in Table~\ref{tab-mtop}
and the boosted measurement is seen to dominate the combination.
The combination is then performed using the two POI measurements and the two measurements of the recoil NP. This yields a combined measurement of $\mt$~with an uncertainty of $0.48$~\GeV,
improving on the combination using only the POIs. The corresponding BLUE weights are shown in Table~\ref{tab-mtop}.
The measurement of the recoil NP from the boosted measurement enters with significant weight, which can be viewed as propagating the information on this NP to the \Jpsi~measurement as anticipated in Section~\ref{sec-nonprof}.
The weight of the $\Jpsi$~analysis is seen to increase when the recoil NP is added to the combination, which is consistent with the \Jpsi~analysis effectively having higher precision once the
information on the recoil NP is available.
Adding the measurements of the remaining NPs into the combination produces a negligible improvement in the precision of $\mt$. This is because the other NPs that are constrained in the boosted measurement
have only a small effect on the $\Jpsi$~measurement.

This combination of the two ATLAS \mt~measurements provides a realistic example in which including the measurements of the NPs in a combination can increase the precision of the POI.
The decision of which NP to include in a real combination of a parameter like \mt~should consider both the potential increase in the precision and the extent to which it is reasonable
to propagate the information on the NPs across different measurements.

\section{Summary}
\label{sec-summary}

In this paper, we consider combinations of measurements that have been performed with full-likelihood or statistical-only likelihood fits.
The work of Ref.~\cite{matrixPLpaper} sets out how to obtain the breakdown of uncertainties for full-likelihood fits and makes their inclusion
 in BLUE combinations straightforward.
 We have shown through two examples that BLUE combinations that use both the parameters of interest (POIs) and the nuisance parameters
 can improve on combinations that use only the POIs.
 In the first example, we also verify with pseudo-experiments that the combined estimates are unbiased with reliable uncertainty estimates.
 The degree of improvement will depend largely on the input measurements and can easily be explored through the software tool that accompanies
 the paper. 
 We have also shown that the Convino tool proposed in Ref.~\cite{Kieseler:2017kxl} can underestimate the uncertainty on a physics parameter when it is
 applied to analyses using full-likelihood fits.
 For cases where it is not possible to make a full joint-likelihood combination, the BLUE method outlined here provides a reliable way to combine measurements
 of physics quantities irrespective of how the original measurements are performed.
 The Combiner open-source software tool~\cite{combiner} provides a full implementation of the method and can be used in future combinations.

\backmatter

\section*{Declarations and acknowledgments}

The likelihoods for the two examples discussed in Section~\ref{sec-example1}~are publicly available in the HS3 format~\cite{hs3}~at~\cite{lhdata}. The software used
to perform the combinations is open-source~\cite{combiner} and the repository includes the configuration files used to produce the results in this article.

For the purpose of open access, the author(s) has applied a Creative Commons Attribution (CC BY) licence to any Author Accepted Manuscript version arising from this submission.
MO acknowledges support from STFC grant ST/W000520/1.

\begin{appendices}

\renewcommand{\thefigure}{\arabic{figure}}
\setcounter{figure}{2}

\section{The minimum of the Convino $\chi^2$}
\label{app-convino}

The first assumption in Ref.~\cite{Kieseler:2017kxl} for the method used in the Convino tool is that the likelihood of a measurement can be approximated as follows\footnote{For brevity, we restrict ourselves here to the Neyman $\chi^2$~and absolute systematic uncertainties.}:
\begin{align}
\chi^2 =& (d_{\mu} - x_{\mu} - k_{\mu i}\lambda_i) M_{\mu \nu} (d_{\nu} - x_{\nu} - k_{\nu j}\lambda_j) \nonumber \\ & +\lambda_i D_{ij} \lambda_j + \lambda_i\lambda_i,
\end{align}
where $x_{\mu}$~are the POIs, $d_{\mu}$~are the estimates of those POIs from this measurement and $\lambda_i$~are the NPs.
It is worth noting that the estimates of the NPs from the measurement do not appear in the approximate likelihood.
The matrices $M$, $D$~and $k$~are
defined in~Ref.~\cite{Kieseler:2017kxl}; the precise definition is not important here, as we only need to know that $M$~and $D$~are symmetric matrices.
The minimum of this $\chi^2$~is found by solving for $\frac{\partial \chi^2}{\partial x_\gamma}=0$~and $\frac{\partial \chi^2}{\partial \lambda_\epsilon}=0$. The first requirement
yields:
\begin{equation}
\frac{\partial \chi^2}{\partial x_\gamma}=-2(d_{\mu} - x_{\mu} - k_{\mu i}\lambda_i)M_{\mu \gamma} =0
\end{equation}
which, for non-zero $M$, has the solution
\begin{equation}
\label{eq-convino-xmin}
(d_{\mu} - x_{\mu} - k_{\mu i}\lambda_i)=0.
\end{equation}
The second requirement yields:
\begin{equation}
\label{eq-convino-npmin}
\frac{\partial \chi^2}{\partial \lambda_\epsilon}=-2(d_{\mu} - x_{\mu} - k_{\mu i}\lambda_i)M_{\mu \epsilon}  + 2D_{\epsilon j}\lambda_j + 2\lambda_{\epsilon}=0.
\end{equation}
Substituting Equation~\ref{eq-convino-xmin} into Equation~\ref{eq-convino-npmin} yields $2D_{\epsilon j}\lambda_j + 2\lambda_{\epsilon}=0$, which has the solution $\lambda_{\epsilon}=0$.
The minimum of the $\chi^2$~is therefore:
\begin{eqnarray}
\lambda_\epsilon &=& 0 \\
x_\mu &=& d_\mu.
\end{eqnarray}
The values of the POIs at the minimum $\chi^2$~correctly correspond to the measured values, but the values of the NPs are always 0.
For a full-likelihood measurement the actual best-fit values of the NPs are not at zero and hence this approximate likelihood is not appropriate in many cases.

The only case in which the approximation appears to be valid is when the measurement has no constraining power on the NPs, in which case the NPs must fit to 0.
As discussed in~Ref.~\cite{Kieseler:2017kxl} the matrix $D$ is then zero and the approximate likelihood simplifies to:
\begin{equation}
\chi^2 = (d_{\mu} - x_{\mu} - k_{\mu i}\lambda_i) M_{\mu \nu} (d_{\nu} - x_{\nu} - k_{\nu j}\lambda_j) + \lambda_i\lambda_i.
\end{equation}
This form of $\chi^2$~has $M$~as the inverse of the statistical covariance matrix, and $k_{\nu j}$~as the impact of systematic source $j$~on parameter estimate $\nu$; it is known~\cite{Valassi:2003mu} to be entirely equivalent to:
\begin{equation}
\chi^2 = (d_{\mu} - x_{\mu} ) C_{\mu \nu} (d_{\nu} - x_{\nu})
\end{equation}
where $C$~is the inverse of the full covariance matrix and the minimum of this $\chi^2$~corresponds to the BLUE combination of the POIs.
The equivalence between Convino and BLUE for analyses without constraints on the NP has been seen numerically in previous results~\cite{lhcmtop,CMS:2021oxn,CMS:2020cso,CMS:2019oeb}.
The results in Section~\ref{sec-example1-convino} show that the Convino tool can underestimate the combined uncertainties on the POIs when combining measurements with constraints on the NPs.
We conclude that the approximation in the Convino tool is only appropriate for measurements
with no constraints on the NPs and it should not be used to combine full-likelihood analyses.

\section{Description of the Combiner configuration files}
\label{app-combiner}

The requirements, installation instructions, and available configuration options for Combiner version \verb|1.0.0| are provided in the dedicated README file included with the software~\cite{combiner}.
This appendix provides a brief overview of the structure and content of the Combiner configuration files used for the studies presented in Sections~\ref{sec-example1}~and \ref{sec-example2}.
The configuration files are provided in the \mycode{configs/paper/} directory included with the Combiner software.
The configuration files are written in the YAML format.
The main configuration file for the example in Section~\ref{sec-example1} is \mycode{configs/paper/simple_example/mainConfig.yml} presented in Figure~\ref{fig-config-ab}.

\begin{figure}[hbp]
\begin{lstlisting}[language=YAML]
general:
  outputPath: output_simple_example/
  debug: info

measurements:
  - name: measurement_A
  - name: measurement_B

measurement_A:
  configFile: configs/paper/simple_example/measurement_A.yml

measurement_B:
  configFile: configs/paper/simple_example/measurement_B.yml
\end{lstlisting}
\caption{Configuration file for combination of measurements $A$~and $B$.\label{fig-config-ab}}
\end{figure}

The configuration contains a general block (under the \mycode{general} key) specifying the output path and verbosity level of the tool,
followed by a list of measurements and their corresponding configuration files.
Each measurement block specifies the path to its individual YAML configuration files that contain the information relevant to that specific measurement.

For measurement $A$, the configuration is specified in the \mycode{configs/paper/simple_example/measurement_A.yml} file and is provided in Figure~\ref{fig-config-a}.
\begin{figure}
\begin{lstlisting}[language=YAML]
pois:
  - name: SigXsecOverSM
    value: 1.0000000001470613

poiCovarianceMatrix:
  parameters: [SigXsecOverSM]
  matrix:
    - row: [0.0160833896815783]

systematics:
  - parameter: alpha_syst1
    impact: [-0.06312441506973665]
    uncertainty: 0.8277319471080937
    value: 4.446548351744184e-08
  - parameter: alpha_syst2
    impact: [-0.04766848007956167]
    uncertainty: 0.9074369519839052
    value: 4.5879398058445986e-08

systematicCorrelations:
  parameters: [alpha_syst1, alpha_syst2]
  matrix:
    - row: [1]
    - row: [-0.30668883034027916,1]
\end{lstlisting}
\caption{Configuration file for measurement $A$.\label{fig-config-a}}
\end{figure}

\begin{figure*}[htbp]
\begin{lstlisting}[language=YAML]
pois:
  - name: SigXsecOverSM
    value: 1.0
systematics:
  - parameter: alpha_syst1
    value: 0.0
  - parameter: alpha_syst2
    value: 0.0
fullCovarianceMatrix:
  parameters: [SigXsecOverSM, alpha_syst1, alpha_syst2]
  matrix:
    - row: [0.08047131692451856]
    - row: [-0.10791173913380991,0.854755955716354]
    - row: [-0.012419032611051813,-0.329982816574247,0.24750728557720753]
\end{lstlisting}
\caption{Configuration file for measurement $B$.\label{fig-config-b}}
\end{figure*}

The \mycode{pois} block specifies the parameters of interest (POIs) and their measured values.
The \mycode{poiCovarianceMatrix} block provides the covariance matrix between the POIs.
Only the diagonal and the bottom left triangle of the covariance matrix need to be specified (the code assumes the matrix is symmetric).
The \mycode{systematics} block lists the NPs (nuisance parameters), the impacts on the POIs (the $\Gamma_{ij}$ parameters for POI $i$ and NP $j$),
the uncertainty (the post-fit uncertainties from the input measurement), and \mycode{value} for the measured values of the NPs.
The \mycode{systematicCorrelations} block defines the correlation matrix between the NPs.
The provided parameters are used in the code to define the full covariance matrix of the measurement as described in Section~\ref{sec-mathSetup}.

For measurement $B$, the configuration is specified in the \mycode{configs/paper/simple_example/measurement_B.yml} file, follows a structure similar to that of measurement $A$, and is presented in Figure~\ref{fig-config-b}.
Instead of providing the POI covariance matrix and the NPs separately, the \mycode{fullCovarianceMatrix} block provides the complete covariance matrix for the measurement.
The two input formats are equivalent and the user can choose the one most convenient for their combination.
NPs with the same name in different measurements are assumed to be correlated across measurements, so in this case \mycode{alpha_syst1} in measurement $A$~is
assumed to be correlated with \mycode{alpha_syst1} in measurement $B$~and \mycode{alpha_syst2} in measurement $A$~is
assumed to be correlated with \mycode{alpha_syst2} in measurement $B$.
Additional correlations between NPs can be specified in the main configuration file under the heading \mycode{byHandNPcorrelations}.

The combination of the measurements $A$~and $B$ including the estimates of the POI and both NPs is performed via \mycode{runCombiner mainConfig.yml}.
The Combiner tool performs the combination and prints a summary of the combined results in the terminal output.
The results of the combination are stored in the specified output path.
Combined central values of all parameters are stored in a text file with their corresponding uncertainties.
The Combiner tool also provides the contribution of each systematic uncertainty to each combined result via standard uncertainty propagation.
Additionally, the full covariance matrix and correlation matrix between all parameters are stored in separate text files.
To visualise the results, several useful plots are automatically generated and saved in the output path.

To perform the combination without explicitly including the estimates of the NPs, one can run \mycode{runCombiner mainConfig.yml npsToFit=none},
where \mycode{npsToFit} is a regular expression to select which NPs should be included in the combination.
The configuration file located at \mycode{configs/paper/simple_example/mainConfig_ASRonly.yml} represents a modified configuration where only the signal region (SR) for the measurement $A$ is considered, as presented in Section~\ref{sec-example1}.

The top quark mass combination, as presented in Section~\ref{sec-example2}, is configured with the \mycode{configs/paper/mt_example/mainMt.yml} file.
The two files \mycode{configs/paper/mt_example/ATLASboosted.yml}~and \mycode{configs/paper/mt_example/ATLASjpsi.yml} contain the inputs for the two measurements
that are combined. As discussed above, systematic uncertainties with the same name between the two measurements are assumed to be correlated.

\end{appendices}

\bibliography{combPaper}


\begin{thebibliography}{26}
\ifx \bisbn   \undefined \def \bisbn  #1{ISBN #1}\fi
\ifx \binits  \undefined \def \binits#1{#1}\fi
\ifx \bauthor  \undefined \def \bauthor#1{#1}\fi
\ifx \batitle  \undefined \def \batitle#1{#1}\fi
\ifx \bjtitle  \undefined \def \bjtitle#1{#1}\fi
\ifx \bvolume  \undefined \def \bvolume#1{\textbf{#1}}\fi
\ifx \byear  \undefined \def \byear#1{#1}\fi
\ifx \bissue  \undefined \def \bissue#1{#1}\fi
\ifx \bfpage  \undefined \def \bfpage#1{#1}\fi
\ifx \blpage  \undefined \def \blpage #1{#1}\fi
\ifx \burl  \undefined \def \burl#1{\textsf{#1}}\fi
\ifx \doiurl  \undefined \def \doiurl#1{\url{https://doi.org/#1}}\fi
\ifx \betal  \undefined \def \betal{\textit{et al.}}\fi
\ifx \binstitute  \undefined \def \binstitute#1{#1}\fi
\ifx \binstitutionaled  \undefined \def \binstitutionaled#1{#1}\fi
\ifx \bctitle  \undefined \def \bctitle#1{#1}\fi
\ifx \beditor  \undefined \def \beditor#1{#1}\fi
\ifx \bpublisher  \undefined \def \bpublisher#1{#1}\fi
\ifx \bbtitle  \undefined \def \bbtitle#1{#1}\fi
\ifx \bedition  \undefined \def \bedition#1{#1}\fi
\ifx \bseriesno  \undefined \def \bseriesno#1{#1}\fi
\ifx \blocation  \undefined \def \blocation#1{#1}\fi
\ifx \bsertitle  \undefined \def \bsertitle#1{#1}\fi
\ifx \bsnm \undefined \def \bsnm#1{#1}\fi
\ifx \bsuffix \undefined \def \bsuffix#1{#1}\fi
\ifx \bparticle \undefined \def \bparticle#1{#1}\fi
\ifx \barticle \undefined \def \barticle#1{#1}\fi
\bibcommenthead
\ifx \bconfdate \undefined \def \bconfdate #1{#1}\fi
\ifx \botherref \undefined \def \botherref #1{#1}\fi
\ifx \url \undefined \def \url#1{\textsf{#1}}\fi
\ifx \bchapter \undefined \def \bchapter#1{#1}\fi
\ifx \bbook \undefined \def \bbook#1{#1}\fi
\ifx \bcomment \undefined \def \bcomment#1{#1}\fi
\ifx \oauthor \undefined \def \oauthor#1{#1}\fi
\ifx \citeauthoryear \undefined \def \citeauthoryear#1{#1}\fi
\ifx \endbibitem  \undefined \def \endbibitem {}\fi
\ifx \bconflocation  \undefined \def \bconflocation#1{#1}\fi
\ifx \arxivurl  \undefined \def \arxivurl#1{\textsf{#1}}\fi
\csname PreBibitemsHook\endcsname

\bibitem[\protect\citeauthoryear{Evans and Bryant}{2008}]{LHC}
\begin{barticle}
\beditor{\bsnm{Evans}, \binits{L.}},
\beditor{\bsnm{Bryant}, \binits{P.}} (eds.):
\batitle{{LHC Machine}}.
\bjtitle{JINST}
\bvolume{3},
\bfpage{08001}
(\byear{2008})
\doiurl{10.1088/1748-0221/3/08/S08001}
\end{barticle}
\endbibitem

\bibitem[\protect\citeauthoryear{Lyons et~al.}{1988}]{Lyons:1988rp}
\begin{barticle}
\bauthor{\bsnm{Lyons}, \binits{L.}},
\bauthor{\bsnm{Gibaut}, \binits{D.}},
\bauthor{\bsnm{Clifford}, \binits{P.}}:
\batitle{{How to Combine Correlated Estimates of a Single Physical Quantity}}.
\bjtitle{Nucl. Instrum. Meth. A}
\bvolume{270},
\bfpage{110}
(\byear{1988})
\doiurl{10.1016/0168-9002(88)90018-6}
\end{barticle}
\endbibitem

\bibitem[\protect\citeauthoryear{Valassi}{2003}]{Valassi:2003mu}
\begin{barticle}
\bauthor{\bsnm{Valassi}, \binits{A.}}:
\batitle{{Combining correlated measurements of several different physical
  quantities}}.
\bjtitle{Nucl. Instrum. Meth. A}
\bvolume{500},
\bfpage{391}--\blpage{405}
(\byear{2003})
\doiurl{10.1016/S0168-9002(03)00329-2}
\end{barticle}
\endbibitem

\bibitem[\protect\citeauthoryear{{ATLAS and CMS
  Collaborations}}{2024}]{lhcmtop}
\begin{barticle}
\bauthor{\bsnm{{ATLAS and CMS Collaborations}}}:
\batitle{{Combination of Measurements of the Top Quark Mass from Data Collected
  by the ATLAS and CMS Experiments at $\sqrt{s}=7$ and 8~TeV}}.
\bjtitle{Phys. Rev. Lett.}
\bvolume{132}(\bissue{26}),
\bfpage{261902}
(\byear{2024})
\doiurl{10.1103/PhysRevLett.132.261902}
{\href{https://arxiv.org/abs/2402.08713}{{arXiv:2402.08713}}}
{[hep-ex]}
\end{barticle}
\endbibitem

\bibitem[\protect\citeauthoryear{{ATLAS and CMS
  Collaborations}}{2020}]{CMS:2020ezf}
\begin{barticle}
\bauthor{\bsnm{{ATLAS and CMS Collaborations}}}:
\batitle{{Combination of the W boson polarization measurements in top quark
  decays using ATLAS and CMS data at $\sqrt{s} =$ 8 TeV}}.
\bjtitle{JHEP}
\bvolume{08}(\bissue{08}),
\bfpage{051}
(\byear{2020})
\doiurl{10.1007/JHEP08(2020)051}
{\href{https://arxiv.org/abs/2005.03799}{{arXiv:2005.03799}}}
{[hep-ex]}
\end{barticle}
\endbibitem

\bibitem[\protect\citeauthoryear{{LHCb Collaboration}}{2020}]{LHCb:2019epo}
\begin{barticle}
\bauthor{\bsnm{{LHCb Collaboration}}}:
\batitle{{Precision measurement of the $\Xi_{cc}^{++}$ mass}}.
\bjtitle{JHEP}
\bvolume{02},
\bfpage{049}
(\byear{2020})
\doiurl{10.1007/JHEP02(2020)049}
{\href{https://arxiv.org/abs/1911.08594}{{arXiv:1911.08594}}}
{[hep-ex]}
\end{barticle}
\endbibitem

\bibitem[\protect\citeauthoryear{{ATLAS and CMS
  Collaborations}}{2019}]{ATLAS:2019hhu}
\begin{barticle}
\bauthor{\bsnm{{ATLAS and CMS Collaborations}}}:
\batitle{{Combinations of single-top-quark production cross-section
  measurements and |f$_{LV}$V$_{tb}$| determinations at $ \sqrt{s} $ = 7 and 8
  TeV with the ATLAS and CMS experiments}}.
\bjtitle{JHEP}
\bvolume{05},
\bfpage{088}
(\byear{2019})
\doiurl{10.1007/JHEP05(2019)088}
{\href{https://arxiv.org/abs/1902.07158}{{arXiv:1902.07158}}}
{[hep-ex]}
\end{barticle}
\endbibitem

\bibitem[\protect\citeauthoryear{van Dyk and Lyons}{2023}]{vanDyk:2023tqz}
\begin{botherref}
\oauthor{\bsnm{Dyk}, \binits{D.}},
\oauthor{\bsnm{Lyons}, \binits{L.}}:
{How to Incorporate Systematic Effects into Parameter Determination}
(2023)
{\href{https://arxiv.org/abs/2306.05271}{{arXiv:2306.05271}}}
{[hep-ex]}
\end{botherref}
\endbibitem

\bibitem[\protect\citeauthoryear{{CMS Collaboration}}{2025}]{CMS:2025kzt}
\begin{barticle}
\bauthor{\bsnm{{CMS Collaboration}}}:
\batitle{{Observation of a pseudoscalar excess at the top quark pair production
  threshold}}.
\bjtitle{Rept. Prog. Phys.}
\bvolume{88}(\bissue{8}),
\bfpage{087801}
(\byear{2025})
\doiurl{10.1088/1361-6633/adf7d3}
{\href{https://arxiv.org/abs/2503.22382}{{arXiv:2503.22382}}}
{[hep-ex]}
\end{barticle}
\endbibitem

\bibitem[\protect\citeauthoryear{{ATLAS Collaboration}}{2026a}]{ATLAS:2026dbe}
\begin{barticle}
\bauthor{\bsnm{{ATLAS Collaboration}}}:
\batitle{{Observation of a cross-section enhancement near the $t\bar{t}$
  production threshold in $\sqrt{s}=13$ TeV $pp$ collisions with the ATLAS
  detector}}.
\bjtitle{Rept. Prog. Phys.}
\bvolume{89}(\bissue{5}),
\bfpage{057801}
(\byear{2026})
\doiurl{10.1088/1361-6633/ae60a0}
{\href{https://arxiv.org/abs/2601.11780}{{arXiv:2601.11780}}}
{[hep-ex]}
\end{barticle}
\endbibitem

\bibitem[\protect\citeauthoryear{{ATLAS Collaboration}}{2026b}]{ATLAS:2026pdi}
\begin{botherref}
\oauthor{\bsnm{{ATLAS Collaboration}}}:
{Combination of Higgs boson measurements at $\sqrt{s} =$ 13 TeV and their
  interpretations by the ATLAS experiment}
(2026)
{\href{https://arxiv.org/abs/2608.07332}{{arXiv:2608.07332}}}
{[hep-ex]}
\end{botherref}
\endbibitem

\bibitem[\protect\citeauthoryear{Kieseler}{2017}]{Kieseler:2017kxl}
\begin{barticle}
\bauthor{\bsnm{Kieseler}, \binits{J.}}:
\batitle{{A method and tool for combining differential or inclusive
  measurements obtained with simultaneously constrained uncertainties}}.
\bjtitle{Eur. Phys. J. C}
\bvolume{77}(\bissue{11}),
\bfpage{792}
(\byear{2017})
\doiurl{10.1140/epjc/s10052-017-5345-0}
{\href{https://arxiv.org/abs/1706.01681}{{arXiv:1706.01681}}}
{[physics.data-an]}
\end{barticle}
\endbibitem

\bibitem[\protect\citeauthoryear{{ATLAS and CMS
  Collaborations}}{2023}]{ATLAS:2022aof}
\begin{barticle}
\bauthor{\bsnm{{ATLAS and CMS Collaborations}}}:
\batitle{{Combination of inclusive top-quark pair production cross-section
  measurements using ATLAS and CMS data at $ \sqrt{s} $ = 7 and 8 TeV}}.
\bjtitle{JHEP}
\bvolume{07},
\bfpage{213}
(\byear{2023})
\doiurl{10.1007/JHEP07(2023)213}
{\href{https://arxiv.org/abs/2205.13830}{{arXiv:2205.13830}}}
{[hep-ex]}
\end{barticle}
\endbibitem

\bibitem[\protect\citeauthoryear{{CMS Collaboration}}{2022}]{CMS:2021oxn}
\begin{barticle}
\bauthor{\bsnm{{CMS Collaboration}}}:
\batitle{{Measurements of the associated production of a W boson and a charm
  quark in proton{\textendash}proton collisions at $\sqrt{s}=8\,\text {TeV}
  $}}.
\bjtitle{Eur. Phys. J. C}
\bvolume{82}(\bissue{12}),
\bfpage{1094}
(\byear{2022})
\doiurl{10.1140/epjc/s10052-022-10897-7}
{\href{https://arxiv.org/abs/2112.00895}{{arXiv:2112.00895}}}
{[hep-ex]}
\end{barticle}
\endbibitem

\bibitem[\protect\citeauthoryear{{CMS Collaboration}}{2021}]{CMS:2020cso}
\begin{barticle}
\bauthor{\bsnm{{CMS Collaboration}}}:
\batitle{{Measurement of differential cross sections for Z bosons produced in
  association with charm jets in pp collisions at $\sqrt{s} =$ 13 TeV}}.
\bjtitle{JHEP}
\bvolume{04},
\bfpage{109}
(\byear{2021})
\doiurl{10.1007/JHEP04(2021)109}
{\href{https://arxiv.org/abs/2012.04119}{{arXiv:2012.04119}}}
{[hep-ex]}
\end{barticle}
\endbibitem

\bibitem[\protect\citeauthoryear{{CMS Collaboration}}{2020}]{CMS:2019oeb}
\begin{barticle}
\bauthor{\bsnm{{CMS Collaboration}}}:
\batitle{{Determination of the strong coupling constant
  $\alpha_{S}(m_\mathrm{Z})$ from measurements of inclusive W$^\pm$ and Z boson
  production cross sections in proton-proton collisions at $ \sqrt{\mathrm{s}}
  $ = 7 and 8 TeV}}.
\bjtitle{JHEP}
\bvolume{06},
\bfpage{018}
(\byear{2020})
\doiurl{10.1007/JHEP06(2020)018}
{\href{https://arxiv.org/abs/1912.04387}{{arXiv:1912.04387}}}
{[hep-ex]}
\end{barticle}
\endbibitem

\bibitem[\protect\citeauthoryear{Pinto et~al.}{2024}]{matrixPLpaper}
\begin{barticle}
\bauthor{\bsnm{Pinto}, \binits{A.}},
\bauthor{\bsnm{Wu}, \binits{Z.}},
\bauthor{\bsnm{Balli}, \binits{F.}},
\bauthor{\bsnm{Berger}, \binits{N.}},
\bauthor{\bsnm{Boonekamp}, \binits{M.}},
\bauthor{\bsnm{Chapon}, \binits{{\'E}.}},
\bauthor{\bsnm{Kawamoto}, \binits{T.}},
\bauthor{\bsnm{Malaescu}, \binits{B.}}:
\batitle{{Uncertainty components in profile likelihood fits}}.
\bjtitle{Eur. Phys. J. C}
\bvolume{84}(\bissue{6}),
\bfpage{593}
(\byear{2024})
\doiurl{10.1140/epjc/s10052-024-12877-5}
{\href{https://arxiv.org/abs/2307.04007}{{arXiv:2307.04007}}}
{[physics.data-an]}
\end{barticle}
\endbibitem

\bibitem[\protect\citeauthoryear{Dado et~al.}{}]{combiner}
\begin{botherref}
\oauthor{\bsnm{Dado}, \binits{T.}},
\oauthor{\bsnm{Owen}, \binits{M.}},
\oauthor{\bsnm{Pinamonti}, \binits{M.}}:
Combiner.
Zenodo.
\doiurl{10.5281/zenodo.12007777}
\end{botherref}
\endbibitem

\bibitem[\protect\citeauthoryear{Cover and Thomas}{2005}]{infoTheoryBook}
\begin{bchapter}
\bauthor{\bsnm{Cover}, \binits{T.M.}},
\bauthor{\bsnm{Thomas}, \binits{J.A.}}:
\bctitle{2}.
\bbtitle{Entropy, Relative Entropy, and Mutual Information},
pp. \bfpage{13}--\blpage{55}.
\bpublisher{John Wiley \& Sons},
\blocation{Hoboken, NJ}
(\byear{2005}).
\doiurl{10.1002/047174882X.ch2}
\end{bchapter}
\endbibitem

\bibitem[\protect\citeauthoryear{Cowan et~al.}{2011}]{Cowan:2010js}
\begin{barticle}
\bauthor{\bsnm{Cowan}, \binits{G.}},
\bauthor{\bsnm{Cranmer}, \binits{K.}},
\bauthor{\bsnm{Gross}, \binits{E.}},
\bauthor{\bsnm{Vitells}, \binits{O.}}:
\batitle{{Asymptotic formulae for likelihood-based tests of new physics}}.
\bjtitle{Eur. Phys. J. C}
\bvolume{71},
\bfpage{1554}
(\byear{2011})
\doiurl{10.1140/epjc/s10052-011-1554-0}
{\href{https://arxiv.org/abs/1007.1727}{{arXiv:1007.1727}}}
{[physics.data-an]}.
\bcomment{[Erratum: Eur.Phys.J.C 73, 2501 (2013)]}
\end{barticle}
\endbibitem

\bibitem[\protect\citeauthoryear{{ATLAS Collaboration}}{2025a}]{boosted}
\begin{barticle}
\bauthor{\bsnm{{ATLAS Collaboration}}}:
\batitle{{Measurement of the top quark mass with the ATLAS detector using
  $t\bar{t}$ events with a high transverse momentum top quark}}.
\bjtitle{Phys. Lett. B}
\bvolume{867},
\bfpage{139608}
(\byear{2025})
\doiurl{10.1016/j.physletb.2025.139608}
{\href{https://arxiv.org/abs/2502.18216}{{arXiv:2502.18216}}}
{[hep-ex]}
\end{barticle}
\endbibitem

\bibitem[\protect\citeauthoryear{{ATLAS Collaboration}}{2025b}]{hepdata.158358}
\begin{botherref}
\oauthor{\bsnm{{ATLAS Collaboration}}}:
{Measurement of the top quark mass with the ATLAS detector using $t\bar{t}$
  events with a high transverse momentum top quark}.
{HEPData (\url{https://doi.org/10.17182/hepdata.158358})}.
\url{https://doi.org/10.17182/hepdata.158358}
(2025)
\end{botherref}
\endbibitem

\bibitem[\protect\citeauthoryear{{ATLAS Collaboration}}{2026a}]{jpsi}
\begin{barticle}
\bauthor{\bsnm{{ATLAS Collaboration}}}:
\batitle{{Measurement of the top-quark mass using decays with a
  J/{\ensuremath{\psi}} meson at $ \sqrt{s}=13 $ TeV with the ATLAS detector}}.
\bjtitle{JHEP}
\bvolume{04},
\bfpage{099}
(\byear{2026})
\doiurl{10.1007/JHEP04(2026)099}
{\href{https://arxiv.org/abs/2511.23091}{{arXiv:2511.23091}}}
{[hep-ex]}
\end{barticle}
\endbibitem

\bibitem[\protect\citeauthoryear{{ATLAS Collaboration}}{2026b}]{hepdata.167264}
\begin{botherref}
\oauthor{\bsnm{{ATLAS Collaboration}}}:
{Measurement of the top-quark mass using decays with a $J/\psi$ meson at
  $\sqrt{s}=$13 TeV with the ATLAS detector}.
{HEPData (\url{https://doi.org/10.17182/hepdata.167264})}.
\url{https://doi.org/10.17182/hepdata.167264}
(2026)
\end{botherref}
\endbibitem

\bibitem[\protect\citeauthoryear{Burgard et~al.}{2026}]{hs3}
\begin{botherref}
\oauthor{\bsnm{Burgard}, \binits{C.}},
\oauthor{\bsnm{Schulz}, \binits{O.}},
\oauthor{\bsnm{Stark}, \binits{G.}},
\oauthor{\bsnm{Rembser}, \binits{J.}},
\oauthor{\bsnm{Cello}, \binits{S.}},
\oauthor{\bsnm{Grunwald}, \binits{C.}}:
{HS3: A Descriptive, Interoperable Serialization Standard for Statistical
  Models in High-Energy Physics}
(2026)
{\href{https://arxiv.org/abs/2606.01760}{{arXiv:2606.01760}}}
{[hep-ex]}
\end{botherref}
\endbibitem

\bibitem[\protect\citeauthoryear{Owen}{}]{lhdata}
\begin{botherref}
\oauthor{\bsnm{Owen}, \binits{M.}}:
{Likelihoods used for testing the combination of measurements that are
  simultaneous fits of parameters of interest and systematic uncertainties}.
Enlighten Research Data, The University of Glasgow.
\doiurl{10.5525/gla.researchdata.2414}
\end{botherref}
\endbibitem

\end{thebibliography}

\end{document}